\documentclass[superscriptaddress,aps,pra,twocolumn,showpacs,nofootinbib,longbibliography,showkeys]{revtex4-2}
\usepackage{slashbox}
\usepackage{amsmath,amssymb,amsthm}
\usepackage{easybmat,comment}
\usepackage[colorlinks=true,citecolor=blue,urlcolor=blue, linkcolor = magenta]{hyperref}
\usepackage[pdftex]{graphicx}
\usepackage{times,txfonts}
\usepackage{braket}
\usepackage{color}
\usepackage{natbib}
\usepackage{subcaption}
\usepackage{ragged2e}
\DeclareCaptionJustification{justified}{\justifying}
 \usepackage[compat=0.4]{yquant}
\newcommand{\be}{\begin{equation}}
	\newcommand{\ee}{\end{equation}}
\newcommand{\ba}{\begin{eqnarray}}
	\newcommand{\ea}{\end{eqnarray}}
\newcommand{\ketbra}[2]{|#1\rangle \langle #2|}

\newcommand{\dket}[1]{|#1\rangle\rangle}
\newcommand{\dbra}[1]{\langle\langle #1|}
\newcommand{\dbraket}[2]{\langle\langle #1 | #2 \rangle\rangle}

\usepackage{comment}
\usepackage{tikz}
\usetikzlibrary{quantikz}
\usepackage{graphicx}

\begin{document}
	
\title{Decohered toric code under quantum damping noise and its mapping to a classical spin model}
\author{Nihar Ranjan Dash\textsuperscript{}}
   \email{dash.1@iitj.ac.in}
   \affiliation{Indian Institute of Technology Jodhpur, Rajasthan 342030, India}
\author{Sanjoy Dutta\textsuperscript{}}
    \email{sanjoy@ppisr.res.in}
    \affiliation{Poornaprajna Institute of Scientific Research (PPISR), Bidalur post, Devanahalli, Bengaluru 562164, India}
\author{R. Srikanth\textsuperscript{}}
   \email{srik@ppisr.res.in}
   \affiliation{Poornaprajna Institute of Scientific Research (PPISR), Bidalur post, Devanahalli, Bengaluru 562164, India}
\author{Subhashish Banerjee}
   \email{subhashish@iitj.ac.in}
   \affiliation{Indian Institute of Technology Jodhpur, Rajasthan 342030, India}




\begin{abstract}
We investigate properties of toric codes under realistic damping error channels, which include squeezing, thermal and non-Markovian effects. First, we map the decohered toric code under the generalized amplitude-damping (GAD) and the squeezed generalized amplitude-damping (SGAD) channels to the statistical-mechanical models using the double Hilbert-space formalism. Second, we map the action of the GAD and SGAD channels on the toric code to stochastic Pauli-type errors via Pauli twirling, yielding asymmetric depolarizing channels, and obtain the logical failure probabilities as a function of temperature and squeezing. In both cases, we relate the channel parameters of the GAD and SGAD channels to the spin-coupling constants of the statistical-mechanical model. 
\end{abstract}
\maketitle


\section{Introduction}
Quantum error correction (QEC) \cite{shor1995scheme, steane1996error, lidarandbrun2013qecbook} schemes are essential for realizing large-scale quantum computation. Among the various methods of QEC, topological codes have emerged as promising due to their robustness against local physical errors \cite{KITAEV20032, bombin2006topological, bombin2010topological, lidarandbrun2013qecbook}. The toric code \cite{kitaev1997quantum, KITAEV20032} is the most extensively studied topological quantum error-correcting code with scalable implementations \cite{bluvstein2022quantum, iqbal2025qutrit,  computing2026quantumerrorcorrectiontoric}. In these codes, physical qubits are arranged on a two-dimensional lattice and rely on local stabilizer measurements. A powerful approach to analyzing the performance of toric codes is to map them to statistical-mechanical spin models. In this mapping, physical errors in the quantum code correspond to disorder in the spin interactions, while error configurations are represented as disorder realization on a lattice \cite{dennis2002topological}.

Initially, Dennis \textit{et al.} introduced the mapping of toric codes under bit-flip errors to the two-dimensional (2D) random-bond Ising model (RBIM) \cite{dennis2002topological}. Later, Bombin \textit{et al.} consider the mapping of the toric codes under depolarizing errors to the 2D random Ashkin-Teller (AT) model, which consists of two coupled Ising systems \cite{ashkin1943statistics}. This mapping was broadly generalized for correlated errors in Ref. \cite{chubb2021statistical}. Also, information-theoretical quantities of the decohered toric code, such as relative entropies and coherent information, can be mapped to the correlation functions of order parameters and domain wall free energy in the spin model \cite{fan2024diagnostics}. In particular, an exact expression for the coherent information of the toric code under decoherence has been studied by mapping it to RBIM \cite{lee2025exact}. 

Previous works on mapping decohered states of toric codes to spin models have studied them under incoherent Pauli-type errors \cite{fan2024diagnostics, lee2023quantum}. However, in realistic quantum devices, errors often contain coherent components arising from control imperfections \cite{bravyi2018correcting, venn2023coherent}, unitary rotation errors \cite{behrends2025statistical}, and amplitude damping (AD) \cite{leung1997approximate, chuang1997bosonic, nielsen2010quantum, Omkar2016two}. Simple probabilistic error models cannot fully capture the impact of coherent errors and may lead to distinct logical failure probabilities. Tools such as Pauli twirling \cite{silva2008scalable, sarvepalli2009asymmetric} convert them into effective incoherent channels, such as those modeled by Pauli channels. Topological codes under AD noise and their Pauli-twirled version have been simulated, and their threshold values have been estimated \cite{tomita2014lowdistance, darmawan2017tensor}. Also under the Choi-Jamiolkowski (CJ) isomorphism \cite{CHOI1975285, jamiolkowski1972linear}, the double Hilbert space formalism \cite{lee2023quantum, lee2025symmetry, bao2026mixed} is used to map quantum states and channels in a larger space formed by the tensor product of a Hilbert space with a copy of itself. This allows us to map quantum channels to operators in the double Hilbert space. 

Recent works have focused on studying the mapping of toric codes under coherent errors, such as unitary rotation errors \cite{behrends2025statistical} and AD noise \cite{lee2025mixed}. Particularly in Ref. \cite{behrends2025statistical}, they have expressed the error amplitudes as RBIM partition functions and have mentioned possible extensions of their work to coherent errors such as amplitude damping and Clifford errors \cite{magesan2013modeling, guti2013approximation}. A method based on doubled Hilbert space formalism is used to connect the mixed state phase of the toric code under AD noise and the AT model \cite{lee2025mixed}. The AD channel is the zero-temperature limit of the more general generalized amplitude damping Channel (GAD) \cite{nielsen2010quantum, banerjee2008geometric, banerjee2008symmetry}, and the squeezed generalized amplitude damping channel (SGAD) \cite{srik2008squeezed, Omkar_2013, jeong2019quantum} further extends GAD by including reservoir squeezing. The complete characterization of GAD and SGAD channels can be obtained through the CJ isomorphism, from which the corresponding Kraus operators are derived \cite{Omkar_2013}. Quantum channels can also be generalized to the non-Markovian regime, where the system retains memory of its past interaction with the environment \cite{banerjee2018open, rivas2014quantum, utagi2018nonmarkovian, sabale2024facets, utagi2020temporal}. Unlike the Markovian case, where the information flow is strictly from the system to the bath, and the decay is exponential, the non-Markovian dynamics allow backflow of information from the environment to the system.

In this work, we study the decohered toric code under GAD  and SGAD channels. We map the toric code under GAD and SGAD to the AT model using the double Hilbert-space formalism and connect the spin-coupling constants to the damping and squeezing parameters. We also extend the former channel to the non-Markovian dynamics regime by incorporating environmental memory, and study its effect on coupling constants. Furthermore, we use the Pauli-twirling tool to map the GAD and SGAD channels into errors in the Pauli basis, yielding asymmetric depolarizing channels. This enables us to investigate the effect of temperature and squeezing on the logical failure probabilities.

The paper is structured as follows. In Sec. \ref{sec:prelim}, we present preliminaries on the mapping of the toric code under decoherence and depolarizing noise to the statistical-mechanical model, and discuss the double Hilbert space formalism and the Pauli-twirling tool. In Secs. \ref{sec:gad} and \ref{sec:sgad}, we study the decohered toric code under GAD and SGAD channels, respectively. The Pauli-twirling tool is used to map both GAD and SGAD channels to the asymmetric depolarizing channels in Sec. \ref{sec:pauli_twirling}. Finally, in Sec. \ref{sec:conclusion} we present our conclusions with discussions.

\section{Preliminaries}\label{sec:prelim}

\subsection{Toric code}
\begin{figure}[t]
    \centering
    \includegraphics[width=5cm]{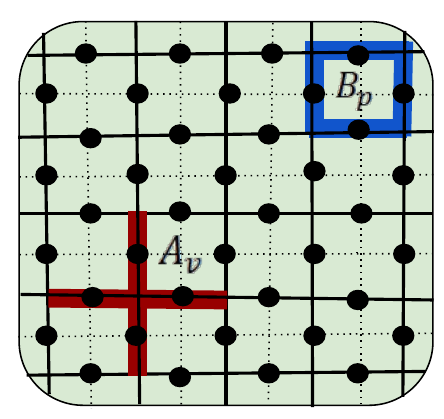}
    \caption{(Color online) The 2D toric code with star and plaquette stabilizers. Black circles embedded on edges represent qubits. On account of the toric topology, we identify the bottommost row (four) qubits with the topmost row qubits, respectively, and the leftmost column qubits with the rightmost column qubits. The primal (resp., dual) lattice is specified by the solid (resp., dotted) connecting lines.}
    \label{fig:toric_code}
\end{figure}

The toric code Hamiltonian on a 2D square lattice $L\times L$ with periodic boundary conditions is given by \cite{KITAEV20032}
\begin{align}
    H&=-\sum_{v}\prod_{e\in {\mathfrak S}_v}X_e -\sum_{p}\prod_{e\in \mathfrak{P}_p}Z_e\\
    &=-\sum_{v}A_v -\sum_{p}B_p,
\end{align}
where $\mathfrak{S}_v$ and $\mathfrak{P}_p$ denote the four vertices belonging to each ``four-pronged star" $v$ and the four vertices belonging to plaquette  $p$ of the lattice, respectively (see Fig. \ref{fig:toric_code}). Here, $A_v$ and $B_p$ are the so-called \textit{star} and \textit{plaquette} stabilizers, respectively.
The toric code on an $L \times L$ square lattice (torus) has
$n = 2L^2$ qubits on edges e, which are constrained by $L^2$ star stabilizers and $L^2$ plaquette stabilizers. However, the stabilizers are not independent and satisfy two global constraints:
$\prod_{\text{all vertices}} A_v = I$ and $\qquad \prod_{\text{all plaquettes}} B_p = I$.
Thus, the number of \emph{independent} stabilizers is $k = 2(L^2 - 1)$. Thus the stabilizer code dimension \cite{gottesman1996class, calderbank1997quantum} is given by 
$
\dim(\text{code space}) = 2^{n-k} = 4.
$
Topologically, this $4$-fold degeneracy corresponds to the $2^{2g}=4$ ground states on a genus $g=1$ surface (the torus). The logical operators are given by non-contractible Wilson loops $W_{1,2}^{z}=\prod_{e\in \ell_{1,2}}Z_e$ and $W_{1,2}^{x}=\prod_{e\in \tilde{\ell}_{1,2}}X_e$ where $\ell_{1,2}$ (resp., $\tilde{\ell}_{1,2}$) forms a closed loop by wrapping around the primal (resp., dual) lattice, which ensures that the logical operators commute with the stabilizers and anti-commute with each other (see Figure \ref{fig:toric_code}). The dual lattice of the square lattice can be understood as a (vertically and horizontally) half a cell shifted copy of the primal lattice, which interchanges star and plaquette stabilizers. On the torus, this shifted lattice wraps around with the same periodic boundary conditions, so the dual lattice is topologically identical to the primal one. Recollect that the code distance is given by the minimum weight of the logical operators. Hence the toric code is a $[[2L^2, 2, L]]$ stabilizer code, able to correct up to $(L-1)/2$ errors.

The ground-state subspace of the toric code is closely connected to the 2D cluster state $\ket{\Psi_{\rm cluster}}$ \cite{briegel2001persistent, Raussendorf01072002, raussendorf2001cluster}. The graph for the cluster state is obtained by assigning qubits to the vertices of the corresponding square lattice. The cluster Hamiltonian has the form $H_{\rm cluster}=\sum_{v}h_v +\sum_e h_e$, where $h_v=-Z_vA_v$ and $h_e=-Z_eX_vX_{v'}$ with edge $e$ connecting two vertices $v$ and $v'$. It follows that the form of the cluster state is given as \cite{ chen2024unconventional, chen2024separability}:
\begin{align}
    \ket{\Psi_{\rm cluster}} \propto \prod_{e}(I+Z_eX_vX_{v'})\ket{\mathfrak{z}_v=1,\mathfrak{x}_e=1},
\end{align}
where $h_v\ket{\mathfrak{z}_v=1,\mathfrak{x}_e=1}=1$ with the product states $\ket{\mathfrak{z}_v=1}=\otimes_{v}\ket{z_v=1}$ and $\ket{\mathfrak{x}_e=1}=\otimes_{e}\ket{x_e=1}$. Here we $z_v$ denotes the Z-basis eigenvalues of all vertex qubits, while $x_e$ denotes the X-basis eigenvalues of all edge qubits. 

Projecting all vertex qubits of the cluster state onto $Z_v=+1$ yields a particular toric-code ground state $\ket{\Psi_0}$, namely the one in the $(W^z_1, W^z_2) = (+1,+1)$ topological sector \cite{chen2024unconventional, lee2025mixed}.
Now we note that:
\begin{subequations}
\begin{align}
    \ket{\Psi_0}&\propto \braket{\mathfrak{z}_v=1|\Psi_{\rm cluster}}
    \label{eq:toric_code_gs-a},\\
    &\propto \bra{\mathfrak{z}_v=1}\prod_{e}(I+Z_eX_vX_{v'})\ket{\mathfrak{z}_v=1,\mathfrak{x}_e=1}\label{eq:toric_code_gs-b}, \\
&=\sum_{\{z_e\}, \{x_v\}}\Big[\prod_{e}(1+z_ex_vx_{v'})\Big]\ket{\mathfrak{z}_e},\label{eq:toric_code_gs-c}
\end{align}
\end{subequations}
where Eq. (\ref{eq:toric_code_gs-c}) is obtained by replacing the operators $Z_e$ and $X_v$ by their eigenvalues $z_e$ and $x_v$ after inserting the resolution of identity in the corresponding basis $\sum_{\mathfrak{z}_e, \mathfrak{x}_v} \ketbra{\mathfrak{z}_e, \mathfrak{x}_v}{\mathfrak{z}_e,\mathfrak{x}_v} $ before the product state. In Eq. (\ref{eq:toric_code_gs-c}), note that any edge configuration in $\ket{\Psi_0}$ makes a non-vanishing contribution to the sum iff it satisfies $z_e=x_vx_{v'}$. Further, along any closed loop any vertex $v$ will be repeated and thus appear an even number of times. Therefore, the product of $z_e$ along closed loops around the lattice is $\prod_{e\in l_{1,2}}z_e=1$. Hence $\ket{\Psi_0}$ is an eigenstate of logical operators $W_{1,2}^{z}=\prod_{e\in l_{1,2}}Z_e$ with eigenvalue $+1$, i.e. $W_{1}^{z}\ket{\Psi_0}=W_{2}^{z}\ket{\Psi_0}=\ket{\Psi_0}$, which can be identified with the $(+,+)$ codeword of the toric code subspace.

\subsection{Toric code under depolarizing error}
Under the depolarizing channel, each qubit in quantum state $\rho$ is subject to the map:
\begin{align}
    \mathcal{E}_e(\rho)=(1-p)\rho+ p_X X\rho X+ p_Y Y\rho Y+ p_ZZ\rho Z,
\end{align}
where the probability of Pauli errors on each edge is $p_\sigma=p/3$, with $\sigma=X,Y,Z$. If $n_{E}$ is the number of edges with errors and $n$ is the total number of edges, then the probability of each of the $\mathfrak{D}(n_E,n_X,n_Y) \equiv \frac{n!}{(n-n_E)!n_X!n_Y!(n_E-n_X-n_Y)!}$ configurations $E$ is
\begin{align}
    p(E)=(1-p)^{n-n_E} \left(\frac{p}{3}\right)^{n_E}.
    \label{eq:prob_error_config}
\end{align}
Note that the total probability of a coarse-grained configuration with $n_E$ errors is $\sum_{n_X=0}^{n_E} \sum_{n_Y=0}^{n_E-n_X}\mathfrak{D}(n_E,n_X,n_Y) \times p(E)$.
For the mapping of the toric code under depolarizing noise to a classical spin model \cite{bombin2012strong}, single qubit bit-flip (X) (resp., phase-flip (Z)) error affects the two adjacent plaquette stabilizers of the primal (resp., dual) lattice, and is correspondingly mapped to a two-spin interaction. The situation is different for $Y$ errors, which simultaneously combine bit-flip and phase-flip errors. Consequently, it affects two neighboring plaquette spins from the primal and dual lattices, and is correspondingly mapped to an effective four-spin interaction. The resulting classical spin structure is precisely described by the random bond Ashkin–Teller model \cite{ashkin1943statistics}, where the Hamiltonian takes the Ising form \cite{bombin2012strong}
\begin{align}
    H=-\sum_e(J_Z s_{i}s_{j}+J_X s_{k}'s_{l}'+J_Ys_{i}s_{j}s_{k}'s_{l}'),
\end{align}
where $s_i$ and $s_k'$ are classical spin variables (not quantum operators) that can take values $\pm1$, and live on the plaquettes of the primal and dual lattice with coupling constants $J_Z$ and $J_X$, respectively. For the mapping of the toric code to a spin model, the required spin coupling constants are defined as \cite{bombin2012strong}
\begin{align}
    J_{\sigma}=-\frac{1}{4\beta}\log{\frac{p_Xp_Yp_Z}{p_{\sigma}^2(1-p)}}, ~~ \sigma=X,Y,Z.
    \label{eq:coupling_const_asym_depo}
\end{align}
The Nishimori line \cite{nishimori1981internal, nishimori2001statistical} is a distinguished line in the temperature-disorder (either $(T,p)$ or $(T,J)$) phase diagram of spin glasses and disordered magnetic systems (such as the random-bond Ising model) where the probability distribution of disorder is mathematically balanced with the thermal distribution, i.e., where the structural disorder (characterized by quenched or frozen disorder) is matched by the thermal disorder. The critical point where the Nishimori line intersects with the phase boundary separating the ordered (ferromagnetic) and disordered (paramagnetic) phases determines the error threshold \cite{lidarandbrun2013qecbook}. Along the Nishimori line, the probability of the error configuration $E$ in Eq. (\ref{eq:prob_error_config}) is proportional to the Boltzmann weight as $p(E) \propto e^{-\beta H(E)}$ with inverse temperature $\beta=1/T$. For the depolarizing channel, the coupling constants become equal, $J_X=J_Y=J_Z\equiv J$, and Eq. (\ref{eq:coupling_const_asym_depo}) reduces to \cite{bombin2010topological, bombin2012strong}
\begin{align}
    J=-\frac{1}{4\beta}\log{\frac{p}{3(1-p)}}.
    \label{eq:coupling_const_depo}
\end{align}
Non-Pauli errors are much harder to map onto a classical spin model with a well-defined Nishimori line, since the computation of the associated effective coupling constants is considerably more cumbersome. In such cases, the double Hilbert Space formalism provides an alternative framework and facilitates a systematic statistical-mechanical analysis.
\subsection{Double Hilbert space formalism}
The double Hilbert space formalism \cite{lee2023quantum, lee2025symmetry, bao2026mixed} provides a way to represent density states and non-unitary dynamics within a unified, state-vector formalism.  As a foundational component of the Choi-Jamiolkowski isomorphism \cite{CHOI1975285, jamiolkowski1972linear}, a density operator $\rho = \sum_{ij}\rho_{ij}\ket{i}\bra{j}$ acting on a $d \times d$ Hilbert space $\mathcal{H}$ is mapped to its $d^2 \times 1$ vectorized form (referred as ``double state") acting on double Hilbert space $\mathcal{H}\otimes \bar{\mathcal{H}}$ defined as $\dket{\rho}=\text{vec}(\rho)\equiv\sum_{ij}\rho_{ij}\dket{i,\bar{j}}$,
where $\dket{i,\bar{j}}\equiv \ket{i}\ket{\bar{j}}$ and the bar corresponds to the copied Hilbert space. Suppose a quantum channel has Kraus operators $\{E_j\}$: $\mathcal{E}(\rho)\equiv \sum_{j}E_j\rho {{E^{\dagger}_j}}$. Using the identity $\text{vec}(AXB)=(A \otimes B^{T})\text{vec}(X)$ for matrices A,X, B, the quantum operation $\mathcal{E}(\rho)$ is mapped to its vectorized form as $\text{vec}(\mathcal{E}(\rho))=(E_j\otimes({E^{\dagger}_j})^{T})\text{vec}(\rho)$ which implies a superoperator $\mathfrak{E}=E_j\otimes({E^{\dagger}_j})^{T}=\sum_{j}E_j\otimes \bar{E_j} $ that acts on $\mathcal{H}\otimes \bar{\mathcal{H}}$. A pure toric code ground state $\rho_0=\ket{\Psi_0}\bra{\Psi_0}$ is mapped to its double state $\dket{\rho_0}=\dket{\Psi_0, \bar{\Psi}_0}$, given by
\begin{align}
 \dket{\rho_0}=\sum_{\mathfrak{z}_e,\bar{\mathfrak{z}}_e}\bigg[\sum_{\mathfrak{x}_v,\bar{\mathfrak{x}}_v}\prod_{e}(1+z_ex_vx_{v'})(1+\bar{z}_e\bar{x}_v\bar{x}_{v'})\bigg]\dket{\mathfrak{z}_e,\bar{\mathfrak{z}}_e}.
\end{align}
The decohered double state becomes $\dket{\rho}=\mathfrak{E}\dket{\rho_{0}}$ and the purity of the $\dket{\rho}$ can computed as $\rm{Tr}(\rho^2)=\dbraket{\rho}{\rho}=\dbra{\rho_{0}}\mathfrak{E}^{\dagger}\mathfrak{E}\dket{\rho_{0}}$ which is mapped to the partition function of a statistical-mechanical model $Z$ as \cite{chen2024unconventional, lee2025mixed}
\begin{align}
    \rm{Tr}(\rho^2)\propto Z\equiv \sum_{\mathfrak{x}_v,\bar{\mathfrak{x}}_v}\sum_{{\mathfrak{t}}_v,\bar{\mathfrak{t}}_v}\prod_{e}\omega_{e},
\end{align}
where
 \begin{align}
 \omega_e&=\sum_{\mathfrak{z}_e,\bar{\mathfrak{z}}_e}\sum_{\mathfrak{z}'_e,\bar{\mathfrak{z}}'_e}\bra{\mathfrak{z}'_e,\bar{\mathfrak{z}}'_e}\mathfrak{E}^{\dagger}\mathfrak{E}\ket{\mathfrak{z}_e,\bar{\mathfrak{z}}_e}
    \times(1+z_ex_vx_{v'})(1+\bar{z}_e\bar{x}_v\bar{x}_{v'})\nonumber\\
    &\times(1+z_e't_v t_{v'})(1+\bar{z}'_e\bar{t}_v\bar{t}_{v'}),
    \label{eq:Boltzmann_weight}
\end{align}
is the Boltzmann weight on edge $e$, and ${x}_v$,$\bar{x}_v$,${t}_v$, and $\bar{t}_v$ are the Ising spins on vertices of the lattice.

\subsection{Pauli twirling}
The Pauli twirling (PT) tool maps an arbitrary quantum channel to a stochastic Pauli channel. The twirling is achieved by averaging the arbitrary noise over the Pauli group, where Kraus operators are represented as probabilistic combinations of Pauli operators \cite{geller2013efficient}. This tool is commonly used in quantum error correction because it enables the analysis of complex noise in terms of Pauli errors, thereby making both analytical and numerical approaches more tractable. The evolution of a density state under a quantum channel is represented as
\begin{align}
\rho\rightarrow\mathcal{E}(\rho)\equiv \sum_{j}E_j\rho {{E_j}^{\dagger}},
\label{eq:evolution_rho}
\end{align} 
where $E_j$ are Kraus operators. Twirling the channel over a set of $K$ operations, $\mathcal{P}=\{P_j\}^{K}_{j=1}$ gives \cite{sarvepalli2009asymmetric, katabarwa2015logical}
\begin{align}
    \mathcal{E}_{\rm twirl}(\rho)=\frac{1}{K}\sum_{j\in1}^{K} P^{\dagger}_{j}\mathcal{E}(P_j\rho P^{\dagger}_{j})P_j.
\end{align}
In PT technique, the twirling operator set is chosen to be the $n$-qubit Pauli basis $\mathcal{P}_n=\{{I,X,Y,Z}\}^{\otimes n}$, comprising a total of $4^n$ distinct elements made of tensor product of Pauli operators $I$, $X$, $Y$, and $Z$. The resulting twirled map can be shown to be diagonal in the Pauli basis, no matter the initial noise $\mathcal{E}$. Thus, compared to the initial channel Eq. (\ref{eq:evolution_rho}), the twirling removes the off-diagonal elements.

\section{Generalized amplitude-damping noise}\label{sec:gad}

\begin{figure}[t]
    \centering
    \includegraphics[width=8cm]{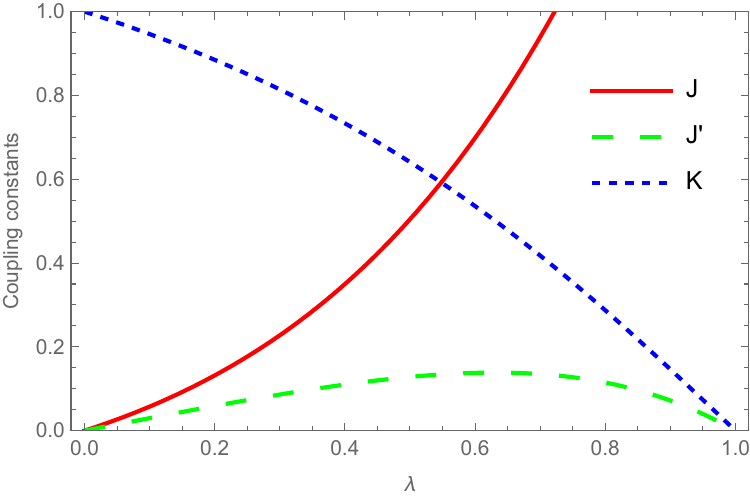}
    \caption{(Color online) Coupling constants for the GAD channel with respect to $\lambda$, with $p=0.8$, whose relations are given in Eq. (\ref{eq:coupli_const_gad}) depicted as the red, green, and blue plots, respectively.}
    \label{fig:coupli_const_gad}
\end{figure}

The generalized amplitude damping channel, which generalizes the AD channel by describing spontaneous emission of a qubit at non-zero temperature \cite{nielsen2010quantum, banerjee2008geometric, banerjee2008symmetry}. In addition to the damping probability $\lambda$, this channel is characterized by an equilibrium excitation probability $p$. It differs from the AD channel by the position of the stationary point of flow on the Bloch sphere. This channel is defined as $\mathfrak{E}_{\rm GAD}\equiv \prod_{e}\mathcal{E}_{e}$, where $\mathcal{E}_{e}(\rho)=\sum_{j=0}^{3} E_{j,e}\rho {E^{\dagger}_{j,e}}$.
The corresponding Kraus operators are given by
\begin{align}
E_{0,e}&=\sqrt{p}\begin{pmatrix}
1 & 0 \\
0 & \sqrt{1-\lambda} 
\end{pmatrix},~~
E_{1,e}=\sqrt{p}\begin{pmatrix}
0 & \sqrt{\lambda} \\
0 & 0 
\end{pmatrix},\nonumber\\
 E_{2,e}&=\sqrt{1-p}\begin{pmatrix}
\sqrt{1-\lambda} & 0 \\
0 & 1 
\end{pmatrix},~~
E_{3,e}=\sqrt{1-p}\begin{pmatrix}
0 & 0 \\
\sqrt{\lambda} & 0 
\end{pmatrix},
\label{eq:kraus_gad}
\end{align}
where $0\leq p \leq 1$ and the damping paramter $\lambda \in[0,1]$ \cite{nielsen2010quantum, banerjee2008geometric, srik2008squeezed}. The decohered toric code state under GAD noise $\rho_{D}=\mathfrak{E}_{\rm GAD}(\rho_{0})$. The purity of the $\rho$ can computed as $Tr(\rho^{2})=\dbra{\rho_{0}}\mathfrak{E}^{\dagger}_{\rm GAD}\mathfrak{E}_{\rm GAD}\dket{\rho_{0}}$, where $\mathfrak{E}_{\rm GAD}=\prod_{e}\mathfrak{E}_{e}$ is given by $\mathfrak{E}_{e}= \sum_{j=0}^{3} E_{j,e}\otimes{\bar{E}_{j,e}}$. In matrix form,

\begin{align}
    \mathfrak{E}_{e}=\begin{pmatrix}
p+(1-p)(1-\lambda) & 0 & 0 & p\lambda\\
0 & \sqrt{1-\lambda} & 0 & 0\\
0 &  0&  \sqrt{1-\lambda} & 0\\
(1-p)\lambda & 0 & 0 & 1-p+p(1-\lambda)
\end{pmatrix}.
\end{align}
\begin{widetext}
Using $\mathfrak{E}_{e}$ we can find that
\begin{align}
&\mathfrak{E}^{\dagger}_{\rm GAD}\mathfrak{E}_{\rm GAD}= \prod_e \mathfrak{E}^{\dagger}_e\mathfrak{E}_e = \prod_{e}\begin{pmatrix}
1-2(1-p)\lambda+2(1-p)^2\lambda^{2} & 0 & 0 & \lambda-2p(1-p)\lambda^{2}\\
0 & 1-\lambda & 0 & 0\\
0 & 0 &  1-\lambda & 0\\
\lambda-2p(1-p)\lambda^{2} & 0 & 0 & p^2\lambda^{2}+(1-p\lambda)^2
\end{pmatrix}.
\end{align}
\end{widetext}
Using Pauli expansion formula $\mathfrak{E}^{\dagger}_{\rm GAD}\mathfrak{E}_{\rm GAD}={\sum_{1,2\in {I,X,Y,Z}}c_{12}}(\sigma_{1}\otimes \sigma_{2})$, we can calculate the coefficients and show that 

\begin{align}
    \mathfrak{E}^{\dagger}_{\rm GAD}\mathfrak{E}_{\rm GAD} &\propto\prod_{e}\Bigg[1+\frac{\lambda(1-\lambda)(2p-1)}{\lambda^{2}(2p^2-2p+1)-2\lambda+2}(Z_e+\bar{Z}_e)\nonumber\\
    &+\frac{1}{\lambda^2(2p^2-2p+1)-2\lambda+2)} \Big\{\lambda^2(2p^2-2p+1)Z_e\bar{Z}_e\nonumber\\
    &+(\lambda-2p(1-p)\lambda^2) X_e \bar{X}_e - (\lambda-2p(1-p)\lambda^2)Y_e\bar{Y}_e \Big\}  \Bigg].
    \label{eq:E_dagger_GAD_E_GAD}
\end{align}
Using Eq. (\ref{eq:E_dagger_GAD_E_GAD}) in Eq. (\ref{eq:Boltzmann_weight}), it can be shown that $\rm{Tr}(\rho^2)$ is mapped to the partition function of the anisotropic AT model as $\rm{Tr}(\rho^2) \propto \sum_{\mathfrak{s}_v,\mathfrak{r}_v}\prod_{e}\omega_{e}$, where the Boltzmann weight $\omega_{e}$ is given by
\begin{align}
\omega_e\propto 1+J s_v s_{v'}+J' r_v r_{v'}+K s_v s_{v'}r_v r_{v'},
\label{eq:spinCC}
\end{align}
where $s_v=x_v\bar{x}_v$ and $r_v=\bar{x}_vt_v$ are the Ising spins on the vertices of the lattice. Extending the formalism of statistical mapping for the decohered toric code under AD channel to the GAD channel \cite{lee2025mixed}, the spin coupling constants for the decohered toric code in Eq. (\ref{eq:spinCC}) are calculated as
\begin{subequations}
\begin{align}
    J&=\frac{\lambda(1+(1-2p)^2\lambda)}{\lambda^2(2p^2-2p+1)-2\lambda+2},
    \label{eq:coupli_const_gad_J}\\
    J'&=\frac{\lambda(1-\lambda)(2p-1)}{\lambda^2(2p^2-2p+1)-2\lambda+2},
    \label{eq:coupli_const_gad_J'}\\
    K&=\frac{\lambda^2-3\lambda+2}{\lambda^2(2p^2-2p+1)-2\lambda+2}.
    \label{eq:coupli_const_gad_K}
    \end{align}
    \label{eq:coupli_const_gad}
\end{subequations}
The two coupling constants $J$ and $K$ show opposite trends with respect to $\lambda$: $J$ increases monotonically and $K$ decreases monotonically (see Fig. \ref{fig:coupli_const_gad}).

The GAD channel can be naturally extended to the non-Markovian dynamics regime by incorporating environmental memory effects. In this extension, the damping parameter becomes time-dependent, and the system evolution is no longer governed by a memoryless exponential decay. The non-Markovian generalized amplitude damping (NMGAD) channel \cite{utagi2024nmgad, wei2025nmgad} is characterized by the Kraus operators
\begin{align}
E_{0,e}&=\sqrt{p}\begin{pmatrix}
1 & 0 \\
0 & \sqrt{1-\lambda(t)} 
\end{pmatrix},~~
E_{1,e}=\sqrt{p}\begin{pmatrix}
0 & \sqrt{\lambda(t)} \\
0 & 0 
\end{pmatrix},\nonumber\\
 E_{2,e}&=\sqrt{1-p}\begin{pmatrix}
\sqrt{1-\lambda(t)} & 0 \\
0 & 1 
\end{pmatrix},~~
E_{3,e}=\sqrt{1-p}\begin{pmatrix}
0 & 0 \\
\sqrt{\lambda(t)} & 0 
\end{pmatrix}.
\label{eq:kraus_NMgad}
\end{align}
Here $0\leq p \leq 1$ and the time-dependent damping paramter $\lambda(t)=1-e^{-gt}(\frac{g}{l} \sinh[\frac{lt}{2}]+\cosh[\frac{lt}{2}]])^{2}$, where $l=\sqrt{g^2-2\lambda_0g}$.

\begin{figure}[t]
    \centering
    \includegraphics[width=8cm]{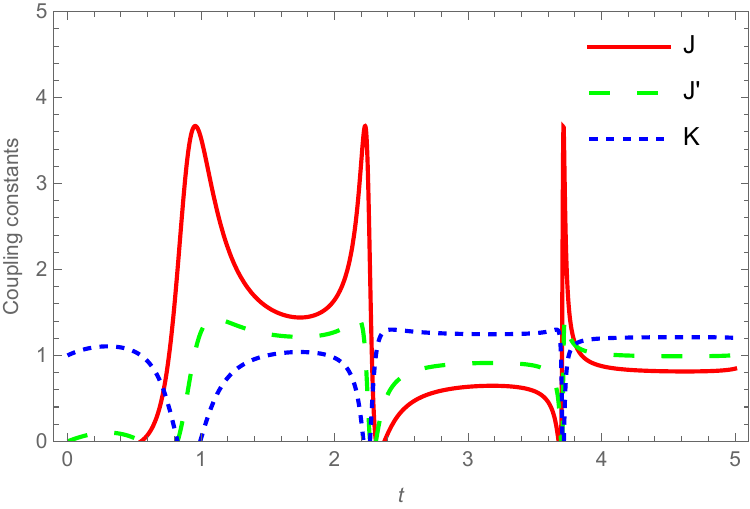}
    \caption{(Color online) Coupling constants for the NMGAD channel with respect to $\lambda$, with $p=0.1$, $g=1$, and $\lambda_0=10$ depicted as the red, green, and blue plots, respectively.}
    \label{fig:coupling_constants_NMGAD}
\end{figure}

Following the same formalism used for the GAD channel, we extend the analysis to the NMGAD case by replacing the constant decay parameter $\lambda$ with a time-dependent function $\lambda(t)$. This simple substitution allows us to retain the structure of the original coupling constants, Eq. (\ref{eq:coupli_const_gad}), while incorporating the environment's memory effects. Upon evaluating and plotting these time-dependent coupling constants, we observe that, unlike the monotonic behavior in the GAD case, the rates now exhibit both increasing and decreasing trends over time (see Fig. \ref{fig:coupling_constants_NMGAD}). Note that for any given time $t$, the coupling constants here will be consistent with the data of Fig. \ref{fig:coupli_const_gad} for the corresponding $\lambda(t)$. This non-monotonic behavior is a signature of non-Markovian dynamics and leads to revivals, indicating a temporary backflow of information from the environment to the system.

\section{Squeezed generalized amplitude-damping noise}\label{sec:sgad}
The SGAD channel is a further extension of the GAD model that describes the interaction of a qubit with a squeezed thermal bath. In addition to the damping parameters, the SGAD channel incorporates squeezing effects described by squeezing parameters. The corresponding Kraus operators are given by \cite{srik2008squeezed, Omkar_2013, jeong2019quantum}
\begin{align}
E_{0,e}&=\sqrt{p}\begin{pmatrix}
1 & 0 \\
0 & \sqrt{1-\alpha} 
\end{pmatrix},~~
E_{1,e}=\sqrt{p}\begin{pmatrix}
0 & \sqrt{\alpha} \\
0 & 0 
\end{pmatrix},\nonumber\\
 E_{2,e}&=\sqrt{1-p}\begin{pmatrix}
\sqrt{1-\nu} & 0 \\
0 &  \sqrt{1-\mu}
\end{pmatrix},~~ \nonumber\\
E_{3,e}&=\sqrt{1-p}\begin{pmatrix}
0 &  \sqrt{\mu}
\sqrt{\nu} & 0 
\end{pmatrix},
\label{eq:Kraus_SGAD}
\end{align}
where the channel parameters are given by
\begin{align}
\nu &= \frac{N}{(1 - p)(2N + 1)} \left(1 - 
\exp\big[-\lambda_0 (2N + 1)t\big]\right), \nonumber\\
\mu &= \frac{2N + 1}{2(1 - p) N} 
\left(
\frac{\sinh^2\left(\frac{a \lambda_0 t}{2}\right)}
{\sinh\left(\frac{\lambda_0 (2N + 1)t}{2}\right)}
\right)
\exp\left[-\frac{\lambda_0 (2N + 1)t}{2}\right], \nonumber\\
\alpha &= \frac{1}{p} \left(
1 - (1 - p)(\mu + \nu) - 
\exp\big[-\lambda_0 (2N + 1)t\big]
\right).
\label{eq:channel_parameter_sgad}
\end{align}
In the Eq. (\ref{eq:channel_parameter_sgad}), $N = N_{\text{th}} \cosh(2r) + \sinh^2(r)$, $N_{\text{th}}=1/(e^{\beta\hbar\omega}-1)$, and $a=\sinh(2r)(2N_{\text{th}}+1)$.

The decohered density state under SGAD noise $\rho_{D}=\mathcal{E}_{\rm SGAD}(\rho_{0})$. The purity of the $\rho_{D}$ can computed as $\text{Tr}(\rho_{D}^{2})=\bra{\rho_{0}}\mathfrak{E}^{\dagger}_{\rm SGAD}\mathfrak{E}_{\rm SGAD}\ket{\rho_{0}}$. Here $\mathfrak{E}_{e}= \sum_{j=0}^{3}{\bar{E}_{j,e}}\otimes E_{j,e}$ and in matrix form,

    \begin{widetext}
\begin{align}
    \mathfrak{E}_{e}=\begin{pmatrix}
p + (1 - p) (1 - \nu) & 0 & 0 & p \alpha + (1 - p)\mu\\
0 & (1-p)\sqrt{1-\mu}\sqrt{1-\nu}+p\sqrt{1-\alpha} & (1-p)\sqrt{\mu}\sqrt{\nu} & 0\\
0 &  (1-p)\sqrt{\mu}\sqrt{\nu}&  (1-p)\sqrt{1-\mu}\sqrt{1-\nu}+p\sqrt{1-\alpha} & 0\\
(1-p)\nu & 0 & 0 & (1-p)(1-\mu)+p(1-\alpha)
\end{pmatrix}.
\label{eq:mathfrak_E_SGAD}
\end{align}
\end{widetext}

Using $\mathfrak{E}_e$ from Eq. (\ref{eq:mathfrak_E_SGAD}) we can find that
\begin{align}
    \mathfrak{E}^{\dagger}_{\rm SGAD}\mathfrak{E}_{\rm SGAD} &\propto\prod_{e}\Big[1+\frac{c_{ZI}}{c_{00}}(Z_e+\bar{Z}_e)\nonumber\\
    &+\frac{1}{c_{00}} \Big\{ c_{ZZ}Z_e \bar{Z}_e +c_{XX} X_e \bar{X}_e +c_{YY} Y_e\bar{Y}_e \Big\}  \Big],
    \label{eq:mathfrak_E_dagger_E_SGAD}
\end{align}
where the coefficients of the Pauli errors are given in Appendix \ref{app_sec:sgad}. Analogous to Eq. \ref{eq:spinCC}, we can write the Boltzmann weight with spin coupling constants that are calculated as 
\begin{subequations}
\begin{align}
J&=\frac{
N_0 - 2p\,N_1 + p^2 N_2
}{
D_0 + p D_1 + p^2 D_2
},
\label{eq:coupli_const_sgad_J}\\
J'&=\frac{
(\mu - \nu) - (\mu^2 - \nu^2) + p N_3 - p^2 N_4
}{
D_3 + p D_4 + p^2 D_5
},
\label{eq:coupli_const_sgad_J'}\\
K&=\frac{
N_5 + 2p\,N_6 + p^2 N_7
}{
D_6 + p D_7 + p^2 D_8
}.
\label{eq:coupli_const_sgad_K}
\end{align}
\label{eq:coupli_const_sgad}
\end{subequations}
where, the numerator ($N_0$ to $N_7$) and denominator ($D_0$ to $D_8$) parameters expression are given in Appendix \ref{app_sec:sgad}.

Within the double Hilbert space formalism, the SGAD channel leads to a corresponding
modification of the effective spin model obtained from the mapping of the decohered toric code.
In particular, the spin model's coupling constants become explicit functions of the squeezing parameters, establishing a direct correspondence between the noise structure and the interaction strengths in the statistical model. This provides a controlled way to study how variations in noise characteristics influence the effective spin interactions and, consequently, the error-correction performance.

\section{Pauli twirling of the GAD and SGAD}\label{sec:pauli_twirling}
\subsection{Pauli-twirled GAD channel}
The Kraus operators of the GAD channel given in Eq. (\ref{eq:kraus_gad}) can be decomposed using Pauli operators as
\begin{align}
    E_{0,e}&=\frac{\sqrt{p}}{2}[(1+\sqrt{1-\lambda})I+(1-\sqrt{1-\lambda})Z],\nonumber\\
     E_{1,e}&=\frac{\sqrt{p}\sqrt{\lambda}}{2}[X+iY],\nonumber\\
    E_{2,e}&=\frac{\sqrt{1-p}}{2}[(1+\sqrt{1-\lambda})I-(1-\sqrt{1-\lambda})Z],\nonumber\\
     E_{3,e}&=\frac{\sqrt{1-p}\sqrt{\lambda}}{2}[X-iY].
    \label{eq:kraus_operators_PT_GAD}
\end{align}
Using Eq. (\ref{eq:kraus_operators_PT_GAD}), $\mathcal{E}_{e}(\rho)=E_{j,e}\rho {E^{\dagger}_{j,e}}$ can be written as
\begin{align}
    \mathcal{E}^{\rm GAD}(\rho)&=\frac{2-2\sqrt{1-\lambda}-\lambda}{4}\rho+\frac{\lambda}{4}X\rho X+\frac{i(1-2p)\lambda}{4}X\rho Y\nonumber\\
    &- \frac{i(1-2p)\lambda}{4}Y\rho X+\frac{\lambda}{4}Y\rho Y-\frac{\lambda}{4}(\rho Z+Z \rho)\nonumber\\
    & + \frac{2-2\sqrt{1-\lambda}-\lambda}{4}Z\rho Z.
\end{align}
\begin{figure}[t]
    \centering
    \includegraphics[width=8cm]{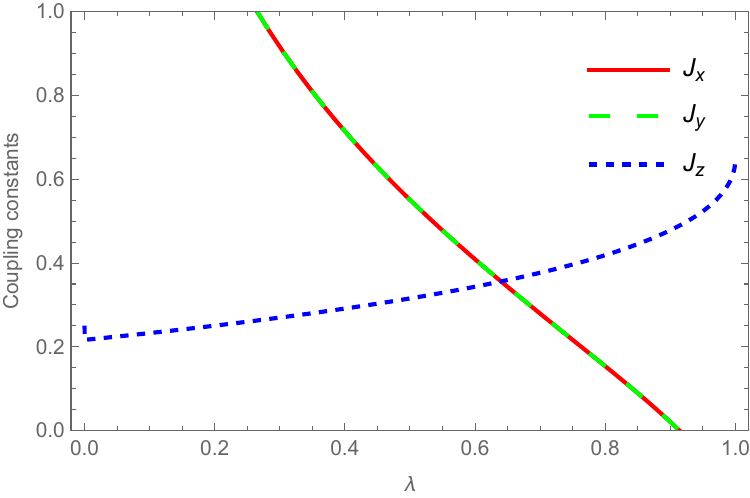}
    \caption{(Color online) Coupling constants for the PT-GAD channel whose relations with $\lambda$ are given in Eq. (\ref{eq:coupling_constants_PTGAD}) depicted as the red, green, and blue plots, respectively.}
    \label{fig:coupling_constants_PT_GAD}
\end{figure}
Under the PT, the off-diagonal terms removed lead to the  PT-GAD channel 
\begin{align}
    \mathcal{E}^{\rm PT-GAD}(\rho)&=\frac{2-2\sqrt{1-\lambda}-\lambda}{4}\rho+\frac{\lambda}{4}X\rho X+\frac{\lambda}{4}Y\rho Y\nonumber\\
    &+ \frac{2-2\sqrt{1-\lambda}-\lambda}{4}Z\rho Z,
\end{align}
with the error probabilities 
\begin{align}
    p_X=p_Y=\frac{\lambda}{4},
    p_Z=\frac{2-2\sqrt{1-\lambda}-\lambda}{4}.
    \label{eq:prob_PTGAD}
\end{align}
Using the values of these probabilities in the Eq.(\ref{eq:coupling_const_asym_depo}), the coupling constants take the forms (see Fig. \ref{fig:coupling_constants_PT_GAD})
\begin{align}
    J_X&=J_Y=-\frac{1}{4\beta}\log{\frac{2-2\sqrt{1-\lambda}-\lambda}{4(1-p)}},\nonumber\\
    &J_Z=-\frac{1}{4\beta}\log{\frac{\lambda^2}{4(1-p)(2-2\sqrt{1-\lambda}-\lambda)}}.
    \label{eq:coupling_constants_PTGAD}
\end{align}
As the decay rate increases, $J_x$ and $J_y$ decrease, indicating that bit-flip errors (associated with $X$ and $Y$) become more probable, thereby weakening the corresponding interactions in the spin model. In contrast, the increase in $J_z$ suggests that phase-flip type errors are relatively less probable, leading to stronger interactions.

\begin{figure}[t]
    \centering
    \includegraphics[width=8cm]{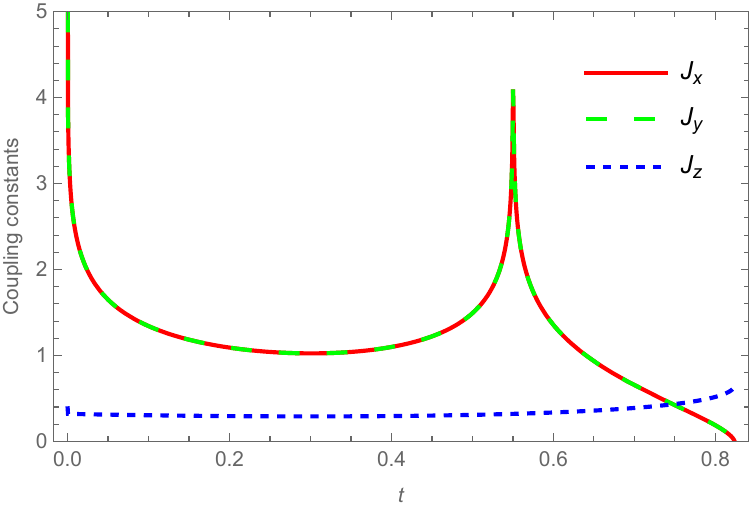}
    \caption{(Color online) Coupling constants for the PT-NMGAD channel with respect to time $t$, with $\beta=1$, $p=0.1$, $g=1$, and $\lambda_0=10$ depicted as the red, green, and blue plots, respectively.}
    \label{fig:coupling_constants_PT_NMGAD}
\end{figure}

By introducing twirling over the NMGAD channel, we obtain a Pauli-twirled NMGAD model in which both the probabilities of Pauli errors and the corresponding coupling constant equations are consistently modified, respectively, in Eqs. (\ref{eq:prob_PTGAD}) and (\ref{eq:coupling_constants_PTGAD}). This is obtained by replacing the damping parameter $\lambda$ with $\lambda(t)$, thereby incorporating environmental memory effects into the framework. We then plot the resulting coupling constant equations for the PT-NMGAD channel (see Fig. \ref{fig:coupling_constants_PT_NMGAD}) and observe a clear signature of non-Markovian behavior. In particular, the coupling constants no longer vary monotonically but instead exhibit alternating regions of increase and decrease, reflecting the underlying information backflow between the system and its environment as discussed earlier in the context of the NMGAD channel.

\subsection{Pauli-twirled SGAD channel}
The Kraus operators of the SGAD channel given in Eq. (\ref{eq:Kraus_SGAD})  can be decomposed using Pauli operators as
\begin{align}
    E_{0,e}&=\frac{\sqrt{p}}{2}[(1+\sqrt{1-\alpha})I+(1-\sqrt{1-\alpha})Z]=aI+bZ,\nonumber\\
     E_{1,e}&=\frac{\sqrt{p}\sqrt{\alpha}}{2}[X+iY],\nonumber\\
    E_{2,e}&=\frac{\sqrt{1-p}}{2}[(\sqrt{1-\nu}+\sqrt{1-\mu})I+(\sqrt{1-\nu}-\sqrt{1-\mu})Z]\nonumber\\
    &=cI+dZ,\nonumber\\
     E_{3,e}&=\frac{\sqrt{1-p}}{2}[(\sqrt{\nu}+\sqrt{\mu})X-i(\sqrt{\nu}-\sqrt{\mu})Y]=eX-ifY.
    \label{eq:Kraus_SGAD_Pauli}
\end{align}
Using Eq. (\ref{eq:Kraus_SGAD_Pauli}), $\mathcal{E}_{e}^{\rm SGAD}(\rho)=E_{j,e}\rho {E^{\dagger}_{j,e}}$ can be written as

\begin{align}
    \mathcal{E}^{\rm SGAD}(\rho)&=(a^2+c^2)\rho+(\frac{p\alpha}{4}+e^2)X\rho X-i(\frac{p\alpha}{4}-ef)X\rho Y\nonumber\\
    &+i(\frac{p\alpha}{4}-ef)Y\rho X+ (\frac{p\alpha}{4}+f^2)Y\rho Y +cd(Z\rho+\rho Z)\nonumber\\
    &+(b^2+d^2)Z\rho Z.
\end{align}

\begin{widetext}
\begin{figure*}[t]
  \begin{subfigure}{.5\textwidth}
  \centering
    \includegraphics[width=1\linewidth]{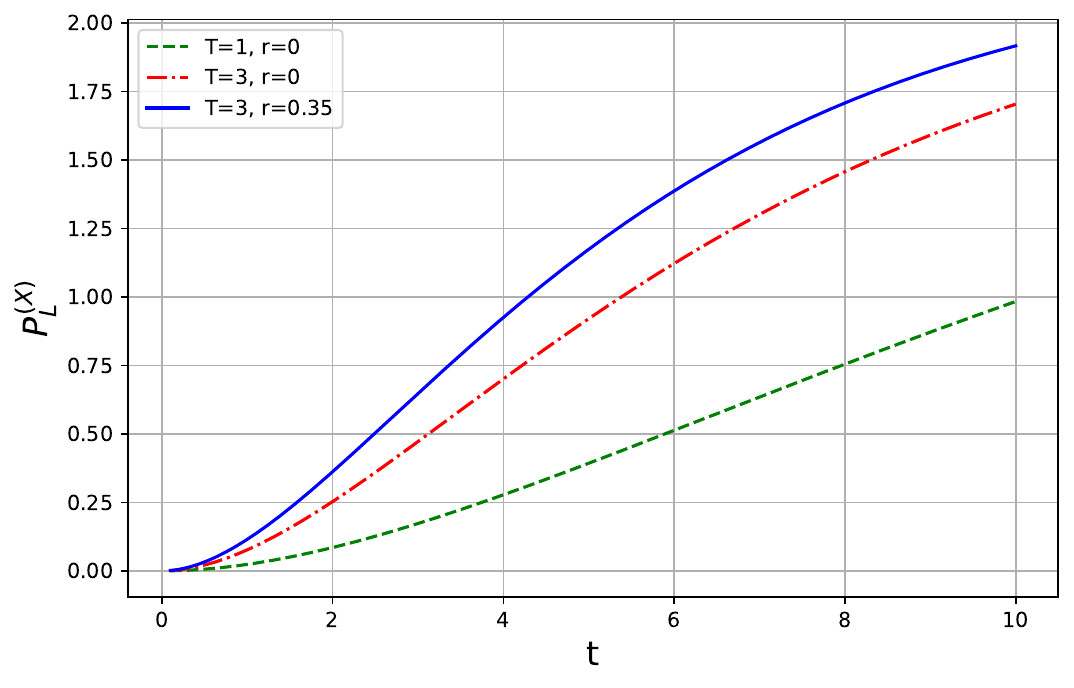}
    \caption{Logical-$X$ failure probability}
  \end{subfigure}%
  \begin{subfigure}{.5\textwidth}
  \centering
    \includegraphics[width=1\linewidth]{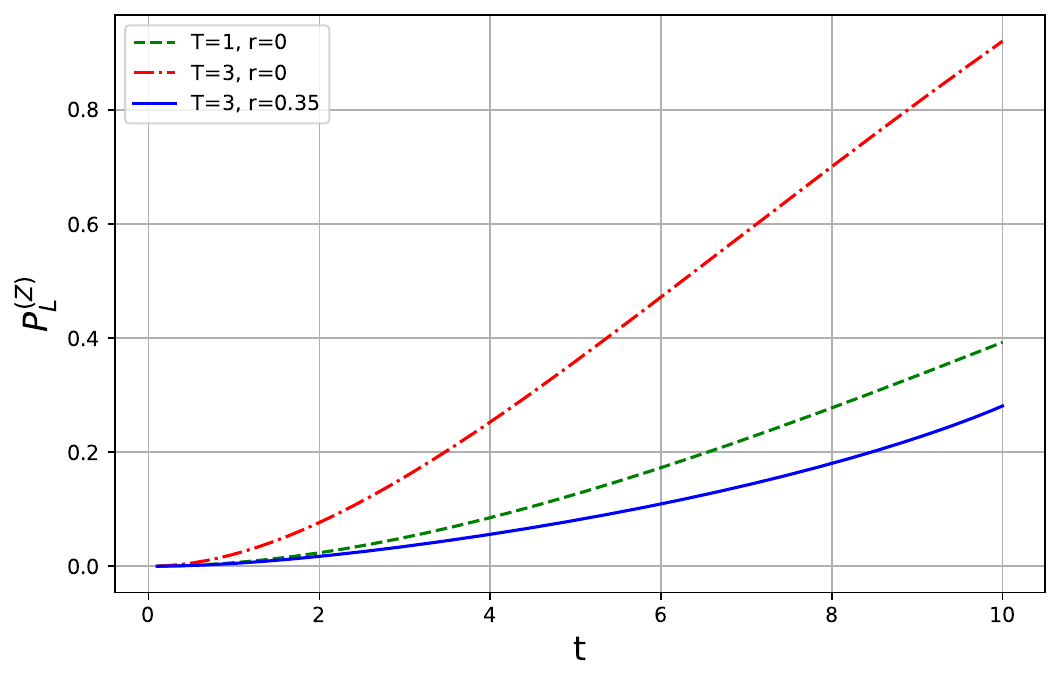}
    \caption{Logical-$Z$ failure probability}
  \end{subfigure}
  \caption{Plot of logical failure probabilities as a function of time $t$. The solid, dashed, and dot-dashed curves correspond to temperature $T=1,3$ and squeezing $r=0,0.35$, respectively.}
  \label{fig:logical_failure_plot}
\end{figure*}
\end{widetext}

Under the PT technique, the off-diagonal terms removed lead to the PT-SGAD channel
\begin{align}
    \mathcal{E}^{\rm PT-SGAD}(\rho)&=(a^2+c^2)\rho+(\frac{p\alpha}{4}+e^2)X\rho X+(\frac{p\alpha}{4}+f^2)Y\rho Y\nonumber\\
    &+ (b^2+d^2)Z\rho Z,
\end{align}
with the error probabilities
\begin{align}
    p_X&=\frac{p\alpha}{4}+e^2\equiv\frac{p\alpha}{4}+\frac{1-p}{4}(\nu+\mu+2\sqrt{\nu\mu}),\nonumber\\
    p_Y&=\frac{p\alpha}{4}+f^2\equiv \frac{p\alpha}{4}+\frac{1-p}{4}(\nu+\mu-2\sqrt{\nu\mu}),\nonumber\\
    p_Z&=b^2+d^2\equiv \frac{p}{4}(2-2\sqrt{1-\alpha}-\alpha)\nonumber\\
    &+\frac{1-p}{4}\Big(2-\nu-\mu-2\sqrt{(1-\nu)(1-\mu)}\Big).
    \label{eq:prob_sgad_twirl}
\end{align}
The bit-flip and phase-flip failure probabilities are given as
\begin{align}
    p_{bf}&=p_X+p_Y,\nonumber\\
    p_{pf}&=p_Y+p_Z.
\end{align}
Using $p_{bf}$ and $p_{pf}$, the logical $X$ and $Z$ failure probabilities of a distance $d$ code are approximately calculated as \cite{ghosh2012surface, fowler2012surface}
\begin{align}P_L^{(X)}&=
d \binom{d}{\frac{d+1}{2}}
\left(p_{bf}\right)^{\frac{d+1}{2}},\nonumber\\
P_L^{(Z)}&=
d \binom{d}{\frac{d+1}{2}}
\left(p_{pf}\right)^{\frac{d+1}{2}}.
\end{align}
The effect of temperature ($T$) and squeezing ($r$) on the logical-$X$ and $Z$ failure probabilities is shown in Fig. \ref{fig:logical_failure_plot}. Increasing $T$ leads to a degradation of both failure probabilities, as evidenced by the difference between the solid and dot-dashed curves. In contrast, increasing $r$ has opposite effects on the two failure probabilities: the $P_L^{(X)}$ degrades, whereas the $P_L^{(Z)}$ improves, as observed from the comparison between the solid and dashed curves. Thus, for error models in which amplitude damping and dephasing are the primary sources of decoherence \cite{sarvepalli2009asymmetric, ghosh2012surface, tomita2014lowdistance}, squeezing can be beneficial to suppress errors. 

Using the values of these probabilities in the Eq. (\ref{eq:coupling_const_asym_depo}), the coupling constants take the forms

\begin{align}
    J_X&=J_Y=-\frac{1}{4\beta}\log \Big[{\frac{p}{4(1-p)}(2-2\sqrt{1-\alpha}-\alpha)}\nonumber\\
    &+\frac{1}{4}\big(2-\nu-\mu-2\sqrt{(1-\nu)(1-\mu)}\big)\Big],\nonumber\\
    &J_Z = -\frac{1}{4\beta} \log \Bigg[
\frac{ \Big(p\alpha + (1-p)(\nu+\mu-2\sqrt{\nu\mu})\Big)^2}{4p(1-p)(2-2\sqrt{1-\alpha}-\alpha)+D}
\Bigg],
\label{eq:coupling_const_sgad_twirl}
\end{align}
where $D=4(1-p)^2\Big(2-\nu-\mu-2\sqrt{(1-\nu)(1-\mu)}\Big)$. Here, both the GAD and SGAD channels are analyzed under the Pauli-twirling method. The twirling method effectively transforms these noise models into equivalent asymmetric depolarizing channels, where the error probabilities along different Pauli directions are no longer identical. This transformation simplifies the analysis while retaining the essential features of the noise.
The resulting asymmetric channels lead to modified spin model representations, where the coupling
constants now reflect the anisotropy introduced by the twirling process. In particular, the effective spin
interactions are determined by the asymmetric error probabilities, which in turn depend on either the
decay rate or the squeezing parameters of the channels.

\section{conclusion}\label{sec:conclusion}
In this work, we investigate toric codes under damping noise by extending known results for the amplitude damping (AD) channel to the generalized amplitude-damping (GAD) channel, in particular considering non-Markovian and squeezed effects. We use the double Hilbert space formalism to map the toric code subjected to these channels to an Ashkin-Teller-type classical spin model and relate the coupling constants to the channel parameters. We also study the implications of the Pauli-twirled version of these channels on the coupling constants under the same mapping. In the case of the Pauli-twirled squeezed GAD channel, we obtain the logical failure probabilities as functions of channel temperature and squeezing, in particular, showing that increasing temperature tends to increase the logical-$X$ and $Z$ failure probabilities, whereas squeezing can improve the logical-$Z$ failure probability.

There are several interesting future directions for studying the toric code suggested by our results. First, to investigate the statistical-mechanical mapping for the damping-induced errors in hybrid bosonic-topological architectures. Recently, this framework has been investigated for concatenating the outer topological code with the inner qubit-to-oscillator code \cite{vuillot2019quantum}. It would be interesting to extend this approach to the topological code that encodes an oscillator into an oscillator grid \cite{barbara2015quantum}. Second, our work can be further extended to investigate the topological code mapping under a combination of noisy channels, such as combined amplitude-damping and dephasing \cite{sarvepalli2009asymmetric, ghosh2012surface, tomita2014lowdistance}. Finally, since biased noise \cite{tuckett2018ultrahigh, huang2023biased} is prevalent across several quantum computing platforms, an interesting direction is to investigate the mapping of topological systems under biased noise and its implications for the error threshold and decoding performance.

\acknowledgments
N.R.D. thanks R.K. Tung for helpful discussions on topological materials. N.R.D. acknowledges financial support from the Department of Science and Technology, Ministry of Science and Technology, India, through the INSPIRE fellowship. 

\appendix
\section{componets of Pauli-twirled SGAD}\label{app_sec:sgad}
\begin{widetext}
The coefficients in Eq. (\ref{eq:channel_parameter_sgad})
\begin{align}
   c_{00}&= \frac{1}{4} \bigg[\left(-1 + 
     p (\alpha - \mu) + \mu]\right)^2 + \left(p (\alpha - \mu) + \
\mu\right)^2 + 
   2 \left(p \sqrt{1 - \alpha]}- ( p-1) \sqrt{1 - \mu]} \sqrt{
       1 - \nu}\right)^2\nonumber\\
       &+ 2 (p-1)^2 \mu \nu + (p-1)^2 {\nu}^2 + \left(1 + (p-1) \nu\right)^2\bigg],\nonumber\\
c_{ZI}=c_{IZ}&=\frac{1}{4} \bigg[\left(-p (1 - \alpha) - (1 - p) (1 - \mu)\right) \left(p (1 - \alpha) + (1 - 
         p) (1 - \mu)\right)+ \left(-p \alpha - (1 - 
         p) \mu\right) \left(p \alpha + (1 - p) \mu\right)\nonumber\\
         &+ \left(-p \sqrt{
       1 - \alpha} - (1 - p) \sqrt{1 - \mu} \sqrt{
       1 - \nu}\right) \left(p \sqrt{1 - \alpha]} + (1 - p) \sqrt{1 - \mu]}
        \sqrt{1 - \nu}\right) + \left(p \sqrt{1 - \alpha} + (1 - p) \sqrt{
      1 - \mu} \sqrt{
      1 - \nu}\right)^2 \nonumber\\
      &+ \left(p + (1 - p) (1 - \nu)\right)^2 + (1 - p)^2 {\nu}^2\bigg],\nonumber\\
      c_{XX}&=  (1 - 
      p) \sqrt{\mu}\sqrt{\nu} \left(p \sqrt{1 - \alpha} - ( p-1) \sqrt{1 - \mu}
        \sqrt{1 - \nu}\right)  +\frac{1}{2} (p-1) \left(-1 + 
      p (\alpha - \mu) + \mu \right) \nu +\frac{1}{2} \Big(p (\alpha - \mu) + \
\mu \Big) \left(1 + ( p-1) \nu\right),\nonumber\\
c_{YY}&=(1 - p) \sqrt{\mu}\sqrt{\nu} \left(p \sqrt{1 - \alpha} - (p-1) \sqrt{1 - \mu}
      \sqrt{1 - \nu}\right)\nonumber\\
      &- 
 \frac{1}{2} (p-1) \left(-1 + p (\alpha - \mu) + \mu\right) \nu - 
 \frac{1}{2} \left(p (\alpha - \mu) + \mu\right) \left(1 + ( p-1) \nu\right).
\end{align}
\end{widetext}

For Eq.(\ref{eq:coupli_const_sgad_J}), the coefficient expressions are given as
\begin{align}
    N_0 &=
\mu + \mu^2 + \nu - 4\mu\nu + \nu^2
+ 2\sqrt{\mu(1-\mu)}\sqrt{\nu(1-\nu)},\nonumber\\
N_1 &=
-1 + \mu^2
+ \sqrt{1-\alpha}\sqrt{1-\mu}\sqrt{1-\nu}
- \sqrt{1-\alpha}\sqrt{\mu}\sqrt{\nu} \nonumber\\
&+ \nu + \alpha\nu + \nu^2
+ 2\sqrt{\mu(1-\mu)}\sqrt{\nu(1-\nu)} - \mu(-1 + \alpha + 4\nu),\nonumber\\
N_2 =&
-2 + \alpha + \alpha^2 + \mu
- 2\alpha\mu + \mu^2 + 2\sqrt{1-\alpha}\sqrt{1-\mu}\sqrt{1-\nu}
\nonumber\\
&- 2\sqrt{1-\alpha}\sqrt{\mu}\sqrt{\nu} + \nu + 2\alpha\nu - 4\mu\nu + \nu^2 \nonumber\\
&+ 2\sqrt{\mu(1-\mu)}\sqrt{\nu(1-\nu)},\nonumber\\
D_0 &= 2 + \mu^2 - 2\mu + 2\mu\nu - 2\nu + \nu^2,\nonumber\\
D_1 &=
-2 + 3\mu - 2\mu^2 + \alpha(-1 + 2\mu) + 2\sqrt{1-\alpha}\sqrt{1-\mu}\sqrt{1-\nu} \nonumber\\
&+ 3\nu - 4\mu\nu - 2\nu^2,\nonumber\\
D_2 &=
2 + \alpha^2 - \mu + \mu^2 - \alpha(1 + 2\mu) - 2\sqrt{1-\alpha}\sqrt{1-\mu}\sqrt{1-\nu} \nonumber\\
&- \nu + 2\mu\nu + \nu^2.
\end{align}

For Eq.(\ref{eq:coupli_const_sgad_J'}), the coefficient expressions are evaluated as
\begin{align}
N_3& = \alpha - \mu - 2\alpha\mu + 2\mu^2 + \nu - 2\nu^2,\nonumber\\
N_4 &= \alpha^2 - 2\alpha\mu + \mu^2 - \nu^2,\nonumber\\
D_3 &= 2 + \mu^2 - 2\mu + 2\mu\nu - 2\nu + \nu^2,\nonumber\\
D_4 &=
-2 + 3\mu - 2\mu^2 + \alpha(-1 + 2\mu)\nonumber\\
&
+ 2\sqrt{1-\alpha}\sqrt{1-\mu}\sqrt{1-\nu}
+ 3\nu - 4\mu\nu - 2\nu^2,\nonumber\\
D_5 &=
2 + \alpha^2 - \mu + \mu^2 - \alpha(1 + 2\mu)\nonumber\\
&
- 2\sqrt{1-\alpha}\sqrt{1-\mu}\sqrt{1-\nu}
- \nu + 2\mu\nu + \nu^2.
\end{align}

For Eq.(\ref{eq:coupli_const_sgad_K}), the coefficient expressions are given as
\begin{align}
N_5 &=
2 + \mu^2 - 3\nu + \nu^2
- 2\sqrt{\mu(1-\mu)}\sqrt{\nu(1-\nu)}
+ \mu(-3 + 4\nu),\nonumber\\
N_6 &=
-1 + 2\mu - \mu^2
+ \sqrt{1-\alpha}\sqrt{1-\mu}\sqrt{1-\nu}
- \sqrt{1-\alpha}\sqrt{\mu}\sqrt{\nu} \nonumber\\
&+ 2\nu - 4\mu\nu - \nu^2
+ 2\sqrt{\mu(1-\mu)}\sqrt{\nu(1-\nu)} + \alpha(-1 + \mu + \nu),\nonumber\\
N_7 &=
2 + \alpha^2 - \mu + \mu^2
- 2\sqrt{1-\alpha}\sqrt{1-\mu}\sqrt{1-\nu} \nonumber\\
&+ 2\sqrt{1-\alpha}\sqrt{\mu}\sqrt{\nu}
- \nu + 4\mu\nu + \nu^2 - 2\sqrt{\mu(1-\mu)}\sqrt{\nu(1-\nu)} \nonumber\\
&- \alpha(1 + 2\mu + 2\nu),\nonumber\\
D_6& = 2 + \mu^2 - 2\mu + 2\mu\nu - 2\nu + \nu^2,\nonumber\\
D_7 &=
-2 + 3\mu - 2\mu^2 + \alpha(-1 + 2\mu) + 2\sqrt{1-\alpha}\sqrt{1-\mu}\sqrt{1-\nu} \nonumber\\
&+ 3\nu - 4\mu\nu - 2\nu^2,\nonumber\\
D_8 &=
2 + \alpha^2 - \mu + \mu^2 - \alpha(1 + 2\mu) - 2\sqrt{1-\alpha}\sqrt{1-\mu}\sqrt{1-\nu} \nonumber\\
&- \nu + 2\mu\nu + \nu^2.
\end{align}

\bibliography{reference}
\end{document}